\documentclass[twocolumn,showpacs,showkeys]{revtex4}
\usepackage{graphicx}
\usepackage{bm}
\usepackage{color}
\usepackage{amsmath}
\usepackage{natbib}

\begin{document}

\title{Kinetic analysis of spin current contribution to spectrum of electromagnetic waves in spin-1/2 plasma, Part I: Dielectric permeability tensor for magnetized plasmas}

\author{Pavel A. Andreev}
\email{andreevpa@physics.msu.ru}
\affiliation{Faculty of physics, Lomonosov Moscow State University, Moscow, Russian Federation.}

 \date{\today}

\begin{abstract}
The dielectric permeability tensor for spin polarized plasmas is derived in terms of the spin-1/2 quantum kinetic model in six-dimensional phase space. Expressions for the distribution function and spin distribution function are derived in linear approximations on the path of dielectric permeability tensor derivation. The dielectric permeability tensor is derived the spin-polarized degenerate electron gas. It is also discussed at the finite temperature regime, where the equilibrium distribution function is presented by the spin-polarized Fermi-Dirac distribution. Consideration of the spin-polarized equilibrium states opens possibilities for the kinetic modeling of the thermal spin current contribution in the plasma dynamics.
\end{abstract}

% PACS, the Physics and Astronomy
                             % Classification Scheme.
%Use showkeys class option if keyword

\maketitle

%52.30.Ex   Two-fluid and multi-fluid plasmas
%52.35.Dm   Sound waves

%52.27.Ep   Electron-positron plasmas

%05.20.Dd	Kinetic theory
%52.25.Dg	Plasma kinetic equations
%52.25.Mq	Dielectric properties
%52.25.Xz	Magnetized plasmas
%52.27.Ny	Relativistic plasmas
%52.27.Gr	Strongly-coupled plasmas
%52.35.Hr	Electromagnetic waves (e.g., electron-cyclotron, Whistler, Bernstein, upper hybrid, lower hybrid)
%75.30.Ds	Spin waves

%%%%%%%%%%TEXT

\section{\label{sec:level1} Introduction}

Spin effects in plasmas were introduced via hydrodynamic formalism \cite{Maksimov VestnMSU 2000, MaksimovTMP 2001, MaksimovTMP 2001 b, Andreev VestnMSU 2007, Marklund PRL07}. Later, a kinetic model was reinstalled in \cite{Brodin PRL 08 g Kin} recapturing results of \cite{Kagan JETP 61 a, Kagan JETP 61 b, Kagan JETP 66}, where the phase space was extended up to eight dimensions to include the spin dependence of the distribution function of particles with fixed module of the spin. Both approaches attract attention of a lot of researchers.  Another kinetic model of spin-1/2 particles was reinstalled in \cite{Oraevsky AP 02, Oraevsky arx, Oraevsky PAN, Andreev kinetics 12, Hurst EPJD 14, Andreev Phys A 15} recapturing results of \cite{Torrey PR 57, Dyson PR 55, Silin JETP 56, Azbel JETP 56, Azbel JETP 57, Azbel JETP 58}, where the distribution functions were considered in the traditional six dimensional phase space. Experimental conformation of results obtained by Dyson \cite{Dyson PR 55} is found in \cite{Feher PR 55}. In this model, the spin evolution appears via presence of the vector distribution function--the spin distribution function in addition to the traditional distribution function. A quantum-relativistic kinetic equations were derived applying the Wigner distribution function \cite{Zhu PPCF 12}. It was derived from a single-particle Dirac equation. Hence, it may neglect relativistic interparticle effects arising between particles of comparable masses \cite{Ivanov PTEP 15}.

Next step in the development of models for the spin-1/2 quantum plasmas was the derivation of the separated spin evolution quantum hydrodynamics \cite{Andreev PRE 15 SEAW}, \cite{Andreev AoP 15 SEAW} (see \cite{Andreev PoP 16 exchange} for a generalized model with exchange interaction) and separated spin evolution quantum kinetics \cite{Andreev PoP 16 sep kin}. In this model the electrons are separated on two subspecies: electrons with spin-up and electrons with spin-down \cite{Andreev PRE 15 SEAW, Harabadze RPJ 04}. The separated spin evolution quantum hydrodynamics is derived from the single particle Pauli equation \cite{Andreev PRE 15 SEAW}. A many-particle derivation was suggested for the separated spin evolution quantum kinetic derivation \cite{Andreev PoP 16 sep kin}. The derivation based on the Pauli equation discover a specific structure of the spin-spin interaction force field. Moreover, it demonstrates unconservation of the particle number in each subspecies. Different behavior of spin-up and spin-down electrons required different pressure. Corresponding equation of state was presented in \cite{Andreev PRE 15 SEAW}. All these features of the separated spin evolution are obtained in Refs. \cite{Andreev PRE 15 SEAW}, \cite{Andreev AoP 15 SEAW} and \cite{Andreev PoP 16 sep kin}, while a two fluid model of electrons was mention in literature earlier \cite{Harabadze RPJ 04}, \cite{Brodin PRL 08 Cl Reg}, \cite{Brodin PRL 10 SPF}. Incompleteness of earlier model bound to incorrect coefficients in spin depending terms \cite{Andreev 1410 Comment}.

Spin evolution leads to the thermal part of the spin flux or the spin current.
The thermal part of the spin current is an analog of the pressure existing in the Euler equation. It is expected that this spin current considerably affects spin properties of plasmas.
However, its analysis requires an equation of state. Corresponding equation of state is found recently with application of the separated spin evolution quantum hydrodynamics \cite{Andreev 1510 Spin Current}:
$$n(\partial_{t}+\textbf{u}\cdot\nabla) \mbox{\boldmath $\mu$}$$
\begin{equation}\label{SC_KA }  -\frac{\hbar}{2m\mu_{e}}\partial^{\beta}[n\mbox{\boldmath $\mu$}\times \partial^{\beta}\mbox{\boldmath $\mu$} ] +\mbox{\boldmath $\Im$}=\frac{2\mu_{e}}{\hbar}n[\mbox{\boldmath $\mu$}\times\textbf{B}], \end{equation}
where $\mbox{\boldmath $\mu$}=\textbf{M}/n$, $\textbf{M}=\textbf{M}(\textbf{r},t)$ is the magnetization of electron gas, $n=n(\textbf{r},t)$ is the concentration of particles, $\textbf{u}(\textbf{r},t)$ is the velocity field, $\textbf{B}$ is the magnetic field, $\hbar$ is the reduced Planck constant, $m$ is the mass of particle, $\mu_{e}$ is the magnetic moment of particle, $\nabla$ and $\partial^{\beta}$ are the vector and tensor notations for the spatial derivatives, and $\mbox{\boldmath $\Im$}$ is the divergence of the thermal part of the spin current, its explicit form is found for the degenerate electron gas \cite{Andreev 1510 Spin Current}:
\begin{equation}\label{SCES spin current many part Vector} \mbox{\boldmath $\Im$}_{P}=\frac{(3\pi^{2})^{2/3}\hbar }{m}(n_{\uparrow}^{2/3}-n_{\downarrow}^{2/3}) [\textbf{M}, \textbf{e}_{z}],\end{equation}
with
$n_{\uparrow}=n-\textrm{M}_{z}/\mid\mu_{e}\mid,$ $n_{\downarrow}=n+\textrm{M}_{z}/\mid\mu_{e}\mid$, where $\mu_{e}$ is the magnetic moment of electron, and subindex $P$ shows that equation (\ref{SCES spin current many part Vector}) is derived from the non-linear Pauli equation, indexes $\uparrow$ and $\downarrow$ refer to the spin-up and spin-down states, correspondingly.
It is an analog of the Fermi pressure. Therefore, it is called the Fermi spin current.
It leads to the modification of spin-plasma wave spectrum \cite{Andreev 1510 Spin Current}. The cut-off frequency of the spin waves is modified by the Fermi spin current. If cut-off frequency of the spin-plasma wave is larger than the plasma frequency the linear interaction of the spin-plasma wave and the ordinary wave appears. Polarization of spin-plasma wave propagating parallel to the external magnetic field changes as well. Hence, a left-hand polarized wave appears instead of the right-hand polarized \cite{Andreev 1510 Spin Current}. Modifications of spectrum and cut-off frequency appear at the perpendicular propagation either.
Similar equation of state was derived from the separated spin evolution quantum kinetics as a moment of the spin distribution function \cite{Andreev PoP 16 sep kin}: $J^{xx}_{K}=J^{yy}_{K}=\cdot3\pi\hbar\mu_{e}(6\pi^{2})^{1/3}(n_{\uparrow}^{4/3}-n_{\downarrow}^{4/3})/32m$ and $J^{xy}_{K}=J^{xz}_{K}=J^{yx}_{K}=J^{yz}_{K}=J^{zx}_{K}=J^{zy}_{K}=J^{zz}_{K}=0$, where subindex $K$ demonstrates that this result is obtained from kinetic model being defined as follows $J^{\alpha\beta}_{K}=\mu_{e}\int S_{0}^{\alpha}(\textbf{p})v^{\beta}dp$. Separate spin evolution affects extraordinary waves directly via the difference of the Fermi pressures for spin-up and spin-down electrons \cite{Andreev 1603}. While, the spin evolution and the Fermi spin current do not affect these waves.

Existence of the Fermi spin current is closely related to difference of the Fermi pressure for the spin-up and spin-down electrons \cite{Andreev 1510 Spin Current}. Difference of the Fermi pressures itself reveals in a modification (an increase) of the pressure contribution in the properties of longitudinal waves. Difference of the Fermi pressures presented in the two fluid model of electron gas leads to extra phenomena: a pair of bulk spin-electron acoustic waves \cite{Andreev PRE 15 SEAW}, \cite{Andreev AoP 15 SEAW}, spin-electron acoustic soliton \cite{Andreev PoP 16 exchange}, surface spin-electron acoustic wave \cite{Andreev APL 16}, a pair of bulk spin-electron-positron acoustic waves in addition to the pair of the bulk spin-electron acoustic waves existing in electron-positron-ion plasmas \cite{Andreev PRE 16}, spin-electron acoustic soliton and spin-electron-positron acoustic soliton in electron-positron-ion plasmas \cite{Andreev_Iqbal PoP 16}. All these waves are longitudinal waves.

The Fermi spin current contributes to the transverse waves \cite{Andreev 1510 Spin Current}.
Results of application of the Fermi spin current require the proper generalization by means of a kinetic model. This paper is devoted to the development of corresponding kinetic model. Presenting derivation of the dielectric permeability tensor for spin polarized plasmas is the first step in this direction.

This paper is organized as follows. In Sec. II, basic nonrelativistic kinetic equations for spin-1/2 plasmas are presented. In Sec. III, linearized kinetic equations and their solutions for the isotropic equilibrium distribution function are obtained. In Sec. IV, the dielectric permeability tensor is found. It is presented in general form and for the spin-polarized Fermi step distribution function.
In Sec. V, a summary of the obtained results is presented.

%\mbox{\boldmath $\mu$}

\section{\label{sec:level1} Quantum kinetic model for spin-1/2 plasmas}

Different quantum kinetic approaches have been developed \cite{Brodin PRL 08 g Kin, Andreev kinetics 12, Hurst EPJD 14, Andreev Phys A 15}, \cite{Andreev PoP 16 sep kin}, \cite{Wigner PR 84, Klimontovich TMP 74, Klimontovich JETP, Klimontovich UFN, Maksimov TMP 2002, Maksimov P D 09, Zamanian EPJD 15}. Some of them do not include the spin evolution \cite{Wigner PR 84, Klimontovich TMP 74, Klimontovich UFN, Maksimov TMP 2002, Maksimov P D 09, Zamanian EPJD 15}, but consider the exchange part of the Coulomb interaction \cite{Klimontovich TMP 74}, \cite{Zamanian EPJD 15}, or consider non-ideal plasmas with strong interaction \cite{Klimontovich JETP}, \cite{Klimontovich UFN}.

The kinetic equation for the distribution function $f(\textbf{r},\textbf{p},t)$ in spin-1/2 plasmas appears as follows \cite{Andreev kinetics 12}, \cite{Andreev Phys A 15}:
$$\partial_{t}f+\textbf{v}\cdot\nabla_{\textbf{r}}f+q_{e}\biggl(\textbf{E}_{ext}+\frac{1}{c}\textbf{v}\times\textbf{B}_{ext}\biggr)\cdot\nabla_{\textbf{p}}f+\mu_{e} (\nabla_{\textbf{r}}^{\beta}B^{\alpha}_{ext})\nabla_{\textbf{p}}^{\beta}S^{\alpha}$$
$$-q_{e}^{2}\int \nabla_{\textbf{r}}G(\textbf{r},\textbf{r}')\cdot\nabla_{\textbf{p}}
f_{2}(\textbf{r},\textbf{p},\textbf{r}',\textbf{p}',t)
d\textbf{r}'d\textbf{p}'$$
\begin{equation}\label{SC_KA kinetic equation gen with spin and int}-\mu_{e}^{2}
\int (\nabla_{\textbf{r}}^{\alpha}G^{\beta\gamma}(\textbf{r},\textbf{r}'))\nabla_{\textbf{p}}^{\alpha} S_{2}^{\beta\gamma}(\textbf{r},\textbf{p},\textbf{r}',\textbf{p}',t) d\textbf{r}'d\textbf{p}'=0, \end{equation}
where $G(\textbf{r},\textbf{r}')=1/|\textbf{r}-\textbf{r}'|$ is the Green function of Coulomb interaction, $G^{\alpha\beta}(\textbf{r},\textbf{r}')=\partial^{\alpha}\partial^{\beta}(1/|\textbf{r}-\textbf{r}'|) +4\pi\delta^{\alpha\beta}\delta(\textbf{r}-\textbf{r}')$ is the Green function of spin-spin interaction, $\nabla_{\textbf{p}}$ is the derivative on the momentum $\textbf{p}=m\textbf{v}$.
It contains the two-particle distribution function $f_{2}(\textbf{r},\textbf{p},\textbf{r}',\textbf{p}',t)$ in the term describing the Coulomb interaction. The spin distribution function $\textbf{S}(\textbf{r},\textbf{p},t)$ arises in the term describing interaction of the spins (the magnetic moments) with the external magnetic field $\textbf{B}_{ext}$. The two-particle spin distribution function $S^{\alpha\beta}(\textbf{r},\textbf{p},\textbf{r}',\textbf{p}',t)$, which is a second rank tensor, appears in the term describing the spin-spin interaction.
The two-particle distribution functions can be reduced to the one-particle distribution functions in the self-consistent (mean-field) approximation. However, equation (\ref{SC_KA kinetic equation gen with spin and int}) requires an additional equation. The equation of evolution of the spin-distribution function $\textbf{S}(\textbf{r},\textbf{p},t)$, which arrears as follows \cite{Andreev kinetics 12}, \cite{Andreev Phys A 15}:
$$\partial_{t}S^{\alpha}+\textbf{v}\cdot\nabla_{\textbf{r}}S^{\alpha} $$ $$+q_{e}\biggl(\textbf{E}_{ext}+\frac{1}{c}\textbf{v}\times\textbf{B}_{ext}\biggr)\cdot\nabla_{\textbf{p}}S^{\alpha}
+\mu_{e} (\nabla_{\textbf{r}}^{\beta} B^{\alpha}_{ext}) \nabla_{\textbf{p}}^{\beta} f$$
$$+q_{e}^{2}
\int \nabla_{\textbf{r}} G(\textbf{r},\textbf{r}')\cdot\nabla_{\textbf{p}} M_{2}^{\alpha}(\textbf{r},\textbf{p},\textbf{r}',\textbf{p}',t) d\textbf{r}'d\textbf{p}'$$
$$-\mu_{e}^{2}
\int (\nabla_{\textbf{r}}^{\gamma}G^{\alpha\beta}(\textbf{r},\textbf{r}'))  \nabla_{\textbf{p}}^{\gamma} N^{\beta}_{2}(\textbf{r},\textbf{p},\textbf{r}',\textbf{p}',t) d\textbf{r}'d\textbf{p}'$$
$$-\frac{2\mu_{e}}{\hbar}\varepsilon^{\alpha\beta\gamma}\Biggl(S^{\beta}(\textbf{r},\textbf{p},t)B^{\gamma}_{ext}(\textbf{r},t)$$
\begin{equation}\label{SC_KA kinetic equation (spin evol) gen with spin and int with two part F} +\mu_{e}\int G^{\gamma\delta}(\textbf{r},\textbf{r}')S_{2}^{\beta\delta}(\textbf{r},\textbf{p},\textbf{r}',\textbf{p}',t)d\textbf{r}'d\textbf{p}'\Biggr)=0.\end{equation}
Equation (\ref{SC_KA kinetic equation (spin evol) gen with spin and int with two part F}) contains the mixed spin-number of particles two-particle distribution function $M_{2}^{\alpha}(\textbf{r},\textbf{p},\textbf{r}',\textbf{p}',t)$ and the number of particles-spin two-particle distribution function $N^{\beta}_{2}(\textbf{r},\textbf{p},\textbf{r}',\textbf{p}',t)$. The quantum terms containing the Planck constant and higher derivatives of the distribution functions are neglected in equations (\ref{SC_KA kinetic equation gen with spin and int}) and (\ref{SC_KA kinetic equation (spin evol) gen with spin and int with two part F}).

The two-particle distribution functions containing in the kinetic equations have the following representation in the self-consistent field approximation
\begin{equation}\label{SC_KA def distribution f2 self consist field}f_{2}(\textbf{r},\textbf{p},\textbf{r}',\textbf{p}',t)=f(\textbf{r},\textbf{p},t) f(\textbf{r}',\textbf{p}',t),\end{equation}
\begin{equation}\label{SC_KA def distribution spin two part function self consist field appr} S_{2}^{\alpha\beta}(\textbf{r},\textbf{p},\textbf{r}',\textbf{p}',t)=S^{\alpha}(\textbf{r},\textbf{p},t)S^{\beta}(\textbf{r}',\textbf{p}',t),\end{equation}
\begin{equation}\label{SC_KA def distribution spin two part function self consist field appr} M_{2}^{\alpha}(\textbf{r},\textbf{p},\textbf{r}',\textbf{p}',t)= S^{\alpha}(\textbf{r},\textbf{p},t)f(\textbf{r}',\textbf{p}',t),\end{equation}
and
\begin{equation}\label{SC_KA def distribution N2 self consist field}N_{2}^{\alpha}(\textbf{r},\textbf{p},\textbf{r}',\textbf{p}',t)=f(\textbf{r},\textbf{p},t) S^{\alpha}(\textbf{r}',\textbf{p}',t).\end{equation}

Equations (\ref{SC_KA kinetic equation gen with spin and int}) and (\ref{SC_KA kinetic equation (spin evol) gen with spin and int with two part F}) are obtained at the neglecting the quantum terms explicitly containing the Planck constant in the kinetic equations (for details see \cite{Andreev kinetics 12}, \cite{Andreev Phys A 15}).

It is assumed in equation (\ref{SC_KA kinetic equation gen with spin and int}) that the kinetic current is reduced to the distribution function $\textbf{J}(\textbf{r}, \textbf{p},t)=\textbf{p} f(\textbf{r}, \textbf{p},t)$ \cite{Andreev kinetics 12} (see equations 12 and 13). Similar approximation is made for the kinetic spin current $J^{\alpha\beta}(\textbf{r}, \textbf{p},t)=p^{\alpha} S^{\beta}(\textbf{r}, \textbf{p},t)$ in the second term in the kinetic equation (\ref{SC_KA kinetic equation (spin evol) gen with spin and int with two part F}) \cite{Andreev kinetics 12} (see equations 36 and 37).

As the result we have next set of equations describing the evolution of spin-1/2 plasma in the self-consistent field approximation \cite{Andreev kinetics 12}, \cite{Hurst EPJD 14}, \cite{Andreev Phys A 15}
\begin{equation}\label{SC_KA kinetic equation gen  classic limit with E and B} \partial_{t}f+\textbf{v}\cdot\nabla_{\textbf{r}}f +q_{e}\biggl(\textbf{E}+\frac{1}{c}\textbf{v}\times\textbf{B}\biggr)\cdot\nabla_{\textbf{p}}f+\nabla^{\alpha}_{\textbf{r}} B^{\beta}\cdot\nabla_{\textbf{p}}^{\alpha} S^{\beta}=0,\end{equation}
and
$$\partial_{t}S^{\alpha}+\textbf{v}\cdot\nabla_{\textbf{r}}S^{\alpha} +q_{e}\biggl(\textbf{E}+\frac{1}{c}\textbf{v}\times\textbf{B}\biggr)\cdot\nabla_{\textbf{p}}S^{\alpha}$$
\begin{equation}\label{SC_KA kinetic equation gen for spin classic limit with E and B}
+\nabla^{\beta}_{\textbf{r}} B^{\alpha}\cdot\nabla_{\textbf{p}}^{\beta}f -\frac{2\mu_{e}}{\hbar}\varepsilon^{\alpha\beta\gamma}S^{\beta}B^{\gamma}=0.\end{equation}

Internal electromagnetic field in equations (\ref{SC_KA kinetic equation gen  classic limit with E and B}) and (\ref{SC_KA kinetic equation gen for spin classic limit with E and B}) appears in the electro-magneto-static limit, the electric field
\begin{equation}\label{SC_KA def of E} \textbf{E}_{int}=q_{e}\int \nabla G(\textbf{r},\textbf{r}') f(\textbf{r}',\textbf{p},t) d\textbf{r}' d\textbf{p} \end{equation}
satisfying quasi-static equations $\nabla\times\textbf{E}=0$ and
\begin{equation}\label{SC_KA } \begin{array}{ccc}\nabla\cdot \textbf{E}=4\pi q_{e}\int f(\textbf{r},\textbf{p},t)d\textbf{p},&   \end{array}\end{equation}
and the magnetic field
\begin{equation}\label{SC_KA def of B} B^{\alpha}_{int}=\mu_{e}\int  G^{\alpha\beta}(\textbf{r},\textbf{r}') S^{\beta}(\textbf{r}',\textbf{p},t) d\textbf{r}' d\textbf{p} \end{equation}
satisfying the following quasi-static equations $\nabla\cdot\textbf{B}=0$ and
\begin{equation}\label{SC_KA rot B with S(r,p,t)} \begin{array}{ccc}\nabla\times \textbf{B}=4\pi\mu_{e}\nabla\times\int \textbf{S}(\textbf{r},\textbf{p},t)d\textbf{p}.&   \end{array}\end{equation}

%\begin{equation}\label{SC_KA } , \end{equation}
To study the plasma dynamics we need to generalize equations (\ref{SC_KA def of E})-(\ref{SC_KA rot B with S(r,p,t)}) arising at the non-relativistic derivation up to the full set of
the Maxwell equations
\begin{equation}\label{SC_KA div E} \begin{array}{ccc}\nabla\cdot \textbf{E}=4\pi\rho,& \nabla\times \textbf{E}=-\frac{1}{c}\partial_{t}\textbf{B},&  \nabla\cdot \textbf{B}=0, \end{array}\end{equation}
%\begin{equation}\label{SC_KA rot E},\end{equation}
%\begin{equation}\label{SC_KA div B}, \end{equation}
and
\begin{equation}\label{SC_KA rot B} \nabla\times \textbf{B}=\frac{1}{c}\partial_{t}\textbf{E}+\frac{4\pi}{c}\textbf{j} +4\pi\nabla\times \textbf{M},\end{equation}
where $\rho=q_{e}\int f(\textbf{r},\textbf{p},t)d\textbf{p}+q_{i}n_{0i}$, $\textbf{j}=q_{e}\int \textbf{v}f(\textbf{r},\textbf{p},t)d\textbf{p}$, $\textbf{M}=\mu_{e}\int \textbf{S}(\textbf{r},\textbf{p},t)d\textbf{p}$ is the magnetization.

Existing kinetic research shows that spin effects leads to damping effects for the electrostatic wave modes \cite{Asenjo PL A 09}, existing of spin waves \cite{Brodin PRL 08 g Kin}, and splitting of Bernstein modes (each mode splits on three branches at the account of the anomalous part of magnetic moment of electrons) \cite{Hussain PP 14 spin bernst}. Kinetic gives an advanced description in compare with hydrodynamics. However, some effects, which do not enter result of traditional hydrodynamics, can be found in extended hydrodynamic models \cite{Andreev IJMP B 15 spin current}, \cite{Zamanian PoP 10}.

\section{Linearized kinetic equations and solutions for distribution functions}

For the derivation of the dielectric permeability tensor we need to consider linear on small perturbations kinetic equations. Assuming that in equilibrium we have non-zero $f_{0}(p)$, $\textbf{B}_{0}=B_{0}\textbf{e}_{z}$, $\textbf{S}_{0}=S_{0}(p)\textbf{e}_{z}$, with $p=|\textbf{p}|$, we find from kinetic equations (\ref{SC_KA kinetic equation gen with spin and int}) and (\ref{SC_KA kinetic equation (spin evol) gen with spin and int with two part F}) the following linearized Fourier transformed kinetic equations
$$-\imath\omega\delta f
+\imath \textbf{v}\cdot\textbf{k}\delta f +\frac{q_{e}}{c}B_{0} (\textbf{v}\times \textbf{e}_{z})\cdot\nabla_{\textbf{p}}\delta f$$
\begin{equation}\label{SC_KA kin eq f lin 3D} +q_{e}\delta \textbf{E}\cdot\nabla_{\textbf{p}}f_{0}+\imath\mu_{e}\delta B_{z}\textbf{k}\cdot\nabla_{\textbf{p}}S_{0}=0,\end{equation}
and
$$-\imath\omega\delta \textbf{S}+\imath (\textbf{v}\cdot\textbf{k}) \delta \textbf{S}
+\frac{q_{e}}{c}B_{0} ((\textbf{v}\times \textbf{e}_{z})\cdot\nabla_{\textbf{p}})\delta \textbf{S}+\imath\mu_{e} (\textbf{k}\cdot\nabla_{\textbf{p}})f_{0}\delta \textbf{B}$$
\begin{equation}\label{SC_KA kin eq S lin 3D} +q_{e}(\delta \textbf{E}\cdot\nabla_{\textbf{p}})\textbf{S}_{0} +\frac{2\mu_{e}}{\hbar}(\textbf{B}_{0}\times\delta \textbf{S}-\textbf{S}_{0}\times \delta \textbf{B})=0,\end{equation}
where the wave vector has the following structure $\textbf{k}=\{k_{x},0,k_{z}\}$ that corresponds to the oblique propagation of waves relatively to the external magnetic field. It corresponds to the isotropic equilibrium distribution functions. More general case of equilibrium distribution function and solutions for linear distribution functions are presented in Appendix A.

As it follows from equation (\ref{SC_KA kin eq S lin 3D}) projections of the spin distribution function $\textbf{S}$ satisfy the
following linear equations
$$\Omega_{e}\partial_{\varphi}\delta S_{x}+\imath (\omega-\textbf{k}\cdot\textbf{v})\delta S_{x}+\Omega_{\mu}\delta S_{y}$$
\begin{equation}\label{SC_KA Sx lin 3D}
=\imath\mu_{e}(\textbf{k}\cdot\nabla_{\textbf{p}})f_{0}\delta
B_{x}+\frac{2\mu_{e}}{\hbar}S_{0z}\delta B_{y},\end{equation} and
$$-\Omega_{\mu}\delta S_{x}+\Omega_{e}\partial_{\varphi}\delta S_{y}+\imath (\omega-\textbf{k}\cdot\textbf{v})\delta S_{y}$$
\begin{equation}\label{SC_KA Sy lin 3D}
=\imath\mu_{e}(\textbf{k}\cdot\nabla_{\textbf{p}})f_{0}\delta
B_{y}-\frac{2\mu_{e}}{\hbar}S_{0}\delta B_{x}, \end{equation}
where $\delta S_{x}$ and $\delta S_{y}$ are bound to each other;
and
$$\Omega_{e}\partial_{\varphi}\delta S_{z}+\imath (\omega-\textbf{k}\cdot\textbf{v})\delta S_{z}$$
\begin{equation}\label{SC_KA Sz lin 3D}=q_{e}(\delta \textbf{E}\cdot\nabla_{\textbf{p}})S_{0z}+\imath\mu_{e}(\textbf{k}\cdot\nabla_{\textbf{p}})f_{0}\delta B_{z}\end{equation}
which is independent from perturbations of other distribution functions. The following notations are used for the charge cyclotron frequency $\Omega_{e}=q_{e}B_{0}/mc$ and the magnetic moment cyclotron frequency $\Omega_{\mu}=2\mu_{e}B_{0}/\hbar$. They are equal to each other if the anomalous part of magnetic moment of electron is neglected. At the transition from equations (\ref{SC_KA kin eq f lin 3D}) and (\ref{SC_KA kin eq S lin 3D}) to equations (\ref{SC_KA Sx lin 3D})-(\ref{SC_KA Sz lin 3D}) we used that $(\textbf{v}\times \textbf{e}_{z})\cdot\partial_{\textbf{p}}=(1/m)\partial_{\varphi}$.

Equation for $\delta f$ (\ref{SC_KA kin eq f lin 3D}) is independent from other equations. Equation for $\delta S_{z}$ (\ref{SC_KA Sz lin 3D}) is independent either. Equations for $\delta S_{x}$ (\ref{SC_KA Sx lin 3D}) and $\delta S_{y}$ (\ref{SC_KA Sy lin 3D}) make a set of equations and should be solved together.

Comparing equation (\ref{SC_KA kin eq f lin 3D}) with spinless case it can be seen that it differs by a single term on the right-hand side. Therefore, solution of equation (\ref{SC_KA kin eq f lin 3D}) can be found in the traditional form
$$\delta f=\frac{1}{\Omega_{e}}\int_{C_{0}}^{\varphi}\biggl(q_{e}(\textbf{v}\cdot\delta \textbf{E})\mid_{\varphi'}\frac{\partial f_{0}}{\partial\varepsilon}+\imath\mu_{e}(\textbf{k}\cdot\textbf{v})\mid_{\varphi'}\delta B_{z}\frac{\partial S_{0z}}{\partial\varepsilon}\biggr)\times $$
\begin{equation}\label{SC_KA f solution}
\times\exp\biggl(\imath\int_{\varphi}^{\varphi'}\frac{1}{\Omega_{e}}(\omega-\textbf{k}\cdot\textbf{v}\mid_{\varphi''})d\varphi''\biggr)d\varphi',\end{equation}
but containing two terms under integral instead of one. In formula (\ref{SC_KA f solution}) and below symbol $\mid_{\varphi'}$ means that it is a function of $\varphi'$. Above, the differentiation on the momentum is replaced by the differentiation on energy $\varepsilon=p^{2}/2m$, $\nabla_{\textbf{p}}=\textbf{v}\partial_{\varepsilon}$.

Equation (\ref{SC_KA Sz lin 3D}) can be solved similarly. As the result, the following solution is found
$$\delta S_{z}=\frac{1}{\Omega_{e}}\int_{C_{3}}^{\varphi}\biggl(q_{e}(\textbf{v}\cdot\delta \textbf{E})\mid_{\varphi'}\frac{\partial S_{0z}}{\partial\varepsilon}+\imath\mu_{e}(\textbf{k}\cdot\textbf{v})\mid_{\varphi'}\delta B_{z}\frac{\partial f_{0}}{\partial\varepsilon}\biggr)\times $$
\begin{equation}\label{SC_KA S z solution}
\times\exp\biggl(\imath\int_{\varphi}^{\varphi'}\frac{1}{\Omega_{e}}(\omega-\textbf{k}\cdot\textbf{v}\mid_{\varphi''})d\varphi''\biggr)d\varphi'.\end{equation}

Spin part of the distribution functions appears via the magnetic field perturbations, which can be represented via the electric field for the further derivation of the dielectric permeability tensor $\delta B_{x}=-k_{z}c\delta E_{y}/\omega$, $\delta B_{y}=c(k_{z}\delta E_{x}-k_{x}\delta E_{z})/\omega$, $\delta B_{z}=k_{x}c\delta E_{y}/\omega$.

Next, the following anzac can be used for simplification of set of equations (\ref{SC_KA Sx lin 3D}) and (\ref{SC_KA Sy lin 3D})
\begin{equation}\label{SC_KA } \delta S_{x}=P(\varphi) \exp\biggl(-\imath\int_{C}^{\varphi}\frac{1}{\Omega_{e}}(\omega-\textbf{k}\cdot\textbf{v}\mid_{\varphi'})d\varphi'\biggr), \end{equation}
and
\begin{equation}\label{SC_KA } \delta S_{y}=R(\varphi) \exp\biggl(-\imath\int_{C}^{\varphi}\frac{1}{\Omega_{e}}(\omega-\textbf{k}\cdot\textbf{v}\mid_{\varphi'})d\varphi'\biggr). \end{equation}
It gives the following equations for functions $P(\varphi)$ and $R(\varphi)$:
$$\Omega_{e}\partial_{\varphi}P(\varphi)+\Omega_{\mu}R(\varphi) =\exp\biggl(\imath\int_{C}^{\varphi}\frac{\omega-\textbf{k}\cdot\textbf{v}\mid_{\varphi'}}{\Omega_{e}}d\varphi'\biggr)\times$$
\begin{equation}\label{SC_KA P lin 3D} \times\biggl(\imath\mu_{e}(\textbf{k}\cdot\nabla_{\textbf{p}})f_{0}\delta B_{x}+\frac{2\mu_{e}}{\hbar}S_{0z}\delta B_{y}\biggr),\end{equation}
and
$$-\Omega_{\mu}P(\varphi)+\Omega_{e}\partial_{\varphi}R(\varphi) =\exp\biggl(\imath\int_{C}^{\varphi}\frac{\omega-\textbf{k}\cdot\textbf{v}\mid_{\varphi'}}{\Omega_{e}}d\varphi'\biggr)\times$$
\begin{equation}\label{SC_KA R lin 3D} \times\biggl(\imath\mu_{e}(\textbf{k}\cdot\nabla_{\textbf{p}})f_{0}\delta B_{y}-\frac{2\mu_{e}}{\hbar}S_{0}\delta B_{x}\biggr).\end{equation}
The structure of functions $P(\varphi)$ and $R(\varphi)$ can be found by solving homogeneous part of equations (\ref{SC_KA P lin 3D}) and (\ref{SC_KA R lin 3D}):
\begin{equation}\label{SC_KA P via  p r} P(\varphi)=\imath p(\varphi) \exp\biggl(\frac{\imath\Omega_{\mu}}{\Omega_{e}}\varphi\biggr)-\imath r(\varphi) \exp\biggl(-\frac{\imath\Omega_{\mu}}{\Omega_{e}}\varphi\biggr), \end{equation}
and
\begin{equation}\label{SC_KA R via  p r} R(\varphi)=p(\varphi) \exp\biggl(\frac{\imath\Omega_{\mu}}{\Omega_{e}}\varphi\biggr) +r(\varphi) \exp\biggl(-\frac{\imath\Omega_{\mu}}{\Omega_{e}}\varphi\biggr). \end{equation}
Equations for $p(\varphi)$ and $r(\varphi)$ can be obtained at the substituting of expressions (\ref{SC_KA P via  p r}) and (\ref{SC_KA R via  p r}) into equations (\ref{SC_KA P lin 3D}) and (\ref{SC_KA R lin 3D}):
\begin{equation}\label{SC_KA p lin 3D} \imath \partial_{\varphi}p(\varphi) \cdot\exp\biggl(\frac{\imath\Omega_{\mu}}{\Omega_{e}}\varphi\biggr)-\imath \partial_{\varphi}r(\varphi) \cdot\exp\biggl(-\frac{\imath\Omega_{\mu}}{\Omega_{e}}\varphi\biggr) =\Pi_{x},\end{equation}
where
$$\Pi_{x}=\frac{1}{\Omega_{e}}\exp\biggl(\imath\int_{C}^{\varphi}\frac{1}{\Omega_{e}}(\omega-\textbf{k}\cdot\textbf{v}\mid_{\varphi'})d\varphi'\biggr)\times$$
\begin{equation}\label{SC_KA Pi x} \times\biggl(\imath\mu_{e}(\textbf{k}\cdot\nabla_{\textbf{p}})f_{0}\delta
B_{x}+\frac{2\mu_{e}}{\hbar}S_{0z}\delta B_{y}\biggr),\end{equation}
and
\begin{equation}\label{SC_KA r lin 3D} \partial_{\varphi}p(\varphi) \cdot\exp\biggl(\frac{\imath\Omega_{\mu}}{\Omega_{e}}\varphi\biggr) +\partial_{\varphi}r(\varphi) \cdot\exp\biggl(-\frac{\imath\Omega_{\mu}}{\Omega_{e}}\varphi\biggr)=\Pi_{y}, \end{equation}
where
$$\Pi_{y}=\frac{1}{\Omega_{e}}\exp\biggl(\imath\int_{C}^{\varphi}\frac{1}{\Omega_{e}}(\omega-\textbf{k}\cdot\textbf{v}\mid_{\varphi'})d\varphi'\biggr)\times$$
\begin{equation}\label{SC_KA Pi y} \times\biggl(\imath\mu_{e}(\textbf{k}\cdot\nabla_{\textbf{p}})f_{0}\delta
B_{y}-\frac{2\mu_{e}}{\hbar}S_{0}\delta B_{x}\biggr).\end{equation}
Independent equations for functions $p(\varphi)$ and $r(\varphi)$
can be found as combinations of equations (\ref{SC_KA p lin 3D})
and (\ref{SC_KA r lin 3D}):
\begin{equation}\label{SC_KA } \partial_{\varphi}p(\varphi)=\frac{1}{2\imath}\exp\biggl(-\frac{\imath\Omega_{\mu}}{\Omega_{e}}\varphi\biggr)(\Pi_{x}+\imath\Pi_{y}), \end{equation}
and
\begin{equation}\label{SC_KA } \partial_{\varphi}r(\varphi)=\frac{1}{2\imath}\exp\biggl(\frac{\imath\Omega_{\mu}}{\Omega_{e}}\varphi\biggr)(-\Pi_{x}+\imath\Pi_{y}). \end{equation}
These equations can be easily integrated. The integration gives the following solutions for functions $p(\varphi)$ and $r(\varphi)$:
\begin{equation}\label{SC_KA } p(\varphi)=\frac{1}{2\imath}\int_{C_{1}}^{\varphi}\exp\biggl(-\frac{\imath\Omega_{\mu}}{\Omega_{e}}\varphi'\biggr)(\Pi_{x}(\varphi')+\imath\Pi_{y}(\varphi')), \end{equation}
and
\begin{equation}\label{SC_KA } r(\varphi)=\frac{1}{2\imath}\int_{C_{2}}^{\varphi}\exp\biggl(\frac{\imath\Omega_{\mu}}{\Omega_{e}}\varphi'\biggr)(-\Pi_{x}(\varphi')+\imath\Pi_{y}(\varphi')). \end{equation}
This calculation leads to the following expressions for the spin distribution functions:
\begin{widetext}$$\delta S_{x}=\frac{\mu_{e}}{2\Omega_{e}}
\Biggl[\int_{C_{1}}^{\varphi}\exp\biggl(\frac{\imath\Omega_{\mu}}{\Omega_{e}}(\varphi-\varphi')\biggr)
\exp\biggl(\imath\int_{\varphi}^{\varphi'}\frac{\omega-\textbf{k}\cdot\textbf{v}\mid_{\varphi''}}{\Omega_{e}}d\varphi''\biggr)
\Biggl((\delta B_{x}+\imath\delta B_{y})(\imath\textbf{k}\cdot\textbf{v}\mid_{\varphi'})\frac{\partial f_{0}}{\partial\varepsilon} +\frac{2S_{0z}}{\hbar}(\delta B_{y}-\imath\delta B_{x})\Biggr)d\varphi'$$
\begin{equation}\label{SC_KA S x solution} +\int_{C_{2}}^{\varphi}\exp\biggl(\frac{\imath\Omega_{\mu}}{\Omega_{e}}(\varphi'-\varphi)\biggr) \exp\biggl(\imath\int_{\varphi}^{\varphi'}\frac{\omega-\textbf{k}\cdot\textbf{v}\mid_{\varphi''}}{\Omega_{e}}d\varphi''\biggr)
\Biggl((\delta B_{x}-\imath\delta B_{y})(\imath\textbf{k}\cdot\textbf{v}\mid_{\varphi'})\frac{\partial f_{0}}{\partial\varepsilon}+\frac{2S_{0z}}{\hbar}(\delta B_{y}+\imath\delta B_{x})\Biggr)d\varphi'\Biggr],\end{equation}
and
$$\delta S_{y}=\frac{\mu_{e}}{2\imath\Omega_{e}}
\Biggl[\int_{C_{1}}^{\varphi}\exp\biggl(\frac{\imath\Omega_{\mu}}{\Omega_{e}}(\varphi-\varphi')\biggr)
\exp\biggl(\imath\int_{\varphi}^{\varphi'}\frac{\omega-\textbf{k}\cdot\textbf{v}\mid_{\varphi''}}{\Omega_{e}}d\varphi''\biggr)
\Biggl((\delta B_{x}+\imath\delta B_{y})(\imath\textbf{k}\cdot\textbf{v}\mid_{\varphi'})\frac{\partial f_{0}}{\partial\varepsilon} +\frac{2S_{0z}}{\hbar}(\delta B_{y}-\imath\delta B_{x})\Biggr)d\varphi'$$
\begin{equation}\label{SC_KA S y solution} +\int_{C_{2}}^{\varphi}\exp\biggl(\frac{\imath\Omega_{\mu}}{\Omega_{e}}(\varphi'-\varphi)\biggr)
\exp\biggl(\imath\int_{\varphi}^{\varphi'}\frac{\omega-\textbf{k}\cdot\textbf{v}\mid_{\varphi''}}{\Omega_{e}}d\varphi''\biggr)
\Biggl((\imath\delta B_{y}-\delta B_{x})(\imath\textbf{k}\cdot\textbf{v}\mid_{\varphi'})\frac{\partial f_{0}}{\partial\varepsilon} -\frac{2S_{0z}}{\hbar}(\delta B_{y}+\imath\delta B_{x})\Biggr)d\varphi'\Biggr].\end{equation}\end{widetext}
Solutions (\ref{SC_KA S x solution}) and (\ref{SC_KA S y solution}) together with solutions (\ref{SC_KA f solution}) and (\ref{SC_KA S z solution}) can be used for derivation of dielectric permeability tensor. Constants $C_{0}$, $C_{1}$, $C_{2}$ and $C_{3}$ are chosen that distribution functions $\delta f$ and $\delta \textbf{S}$ are periodic functions of angle $\varphi$: $\delta f(\varphi+2\pi)=\delta f(\varphi)$ and $\delta \textbf{S}(\varphi+2\pi)=\delta \textbf{S}(\varphi)$, so $C_{i}=\infty$, where $i=0,1,2,3$.

\section{\label{sec:level1} Dielectric permeability tensor for magnetized spin-1/2 plasmas}

After Fourier transformation of the Maxwell equations, the magnetic field taken from equation $\textbf{k}\times\delta\textbf{E}=\omega\delta\textbf{B}/c$ is substituted into equation (\ref{SC_KA rot B}). It leads to
\begin{equation}\label{SC_KA } \biggl[k^{2}\delta^{\alpha\beta}-k^{\alpha}k^{\beta}-\frac{\omega^{2}}{c^{2}}\varepsilon^{\alpha\beta}(\omega)\biggr]\delta E_{\beta}=0, \end{equation}
where the dielectric permeability tensor appears as
$$\varepsilon^{\alpha\beta}(\omega)\delta E_{\beta}=\delta^{\alpha\beta}\delta E_{\beta}+\frac{4\pi\imath}{\omega}\frac{q_{e}}{m}\int p^{\alpha}\delta f d\textbf{p}$$
\begin{equation}\label{SC_KA }-\frac{4\pi\mu_{e}c}{\omega}\int\varepsilon^{\alpha\beta\gamma}k^{\beta}\delta S^{\gamma}d\textbf{p},\end{equation}
and $\delta^{\alpha\beta}$ is the Kronecker symbol.
For further analysis it is useful to distinguish a part of conductivity tensor caused by the current
\begin{equation}\label{SC_KA sigma 1 short}\sigma^{\alpha\beta}_{1}(\omega)\delta E_{\beta}=\frac{q_{e}}{m}\int p^{\alpha}\delta f d\textbf{p}\end{equation}
and another part caused by the curl of magnetization
\begin{equation}\label{SC_KA sigma 2 short}\sigma^{\alpha\beta}_{2}(\omega)\delta E_{\beta}=\imath\mu_{e}c\int\varepsilon^{\alpha\beta\gamma}k^{\beta}\delta S^{\gamma}d\textbf{p}.\end{equation}
The standard resolution between $\varepsilon^{\alpha\beta}$ and $\sigma^{\alpha\beta}$ is used: $\varepsilon^{\alpha\beta}=\delta^{\alpha\beta}+(4\pi\imath/\omega)\sigma^{\alpha\beta}$.

\subsection{General form of dielectric permeability tensor for isotropic distribution functions}

An explicit form of the conductivity tensor (\ref{SC_KA sigma 1 short}), (\ref{SC_KA sigma 2 short}) is presented in Appendix B.
After the integration on $\varphi$ and $\varphi'$ in equations (\ref{SC_KA sigma 1 expl})-(\ref{SC_KA sigma z expl}) (which are presented in Appendix B) we find the following dielectric permeability tensor
$$\varepsilon^{\alpha\beta}=\delta^{\alpha\beta}+\int d\textbf{p}\sum_{n=-\infty}^{\infty}\frac{\Lambda^{\alpha\beta}(n)}{\omega-k_{z}v_{z}-n\Omega_{e}} $$ $$+\sum_{r=+,-}\int d\textbf{p}\sum_{n=-\infty}^{\infty}\frac{\Lambda^{\alpha\beta}_{S,r}(n)}{\omega-k_{z}v_{z}-n\Omega_{e}+r\Omega_{\mu}} $$
\begin{equation}\label{SC_KA dielectric permeability tensor GF is}-4\pi\mu_{e}^{2}\frac{k_{x}^{2}c^{2}}{\omega^{2}}\int d\textbf{p}\sum_{n=-\infty}^{\infty}J_{n}^{2}\frac{\partial f_{0}}{\partial\varepsilon}\delta^{\alpha y}\delta^{\beta y}+\Delta^{\alpha}\delta^{\beta y},\end{equation}
where
$$\Lambda^{\alpha\beta}(n)=\frac{4\pi q_{e}^{2}}{\omega}\frac{\partial f_{0}}{\partial\varepsilon}\Pi^{\alpha\beta}_{Cl}(n)$$
\begin{equation}\label{SC_KA }  +4\pi\mu_{e}^{2}\frac{k_{x}^{2}c^{2}}{\omega}J_{n}^{2}\frac{\partial f_{0}}{\partial\varepsilon}\delta^{\alpha y}\delta^{\beta y} +\frac{4\pi q_{e}\mu_{e}c}{\omega}\Pi^{\alpha\beta}_{S}(n)\frac{\partial S_{0z}}{\partial\varepsilon} ,\end{equation}
with
\begin{equation}\label{SC_KA Pi Cl} \widehat{\Pi}_{Cl}(n)=\left(\begin{array}{ccc}
\frac{\Omega_{e}^{2}}{k_{x}^{2}}n^{2} J_{n}^{2} & \imath v_{\perp}\frac{\Omega_{e}}{k_{x}} nJ_{n}J_{n}' & v_{z}\frac{\Omega_{e}}{k_{x}}n J_{n}^{2} \\
-\imath v_{\perp}\frac{\Omega_{e}}{k_{x}}n J_{n}J_{n}' & v_{\perp}^{2}(J_{n}')^{2} & -\imath v_{\perp}v_{z} J_{n}J_{n}' \\
v_{z} \frac{\Omega_{e}}{k_{x}}n J_{n}^{2} & \imath v_{\perp}v_{z} J_{n}J_{n}' & v_{z}^{2}J_{n}^{2}
\end{array}\right)\end{equation}
describing the classical part and presented in many textbooks (see for instance \cite{Landau v10}, \cite{Rukhadze book 84}),
and the tensor
\begin{equation}\label{SC_KA } \Pi^{\alpha\beta}_{S}(n)=\left(\begin{array}{ccc}
0 & \imath\Omega_{e}nJ_{n}^{2} & 0 \\
-\imath\Omega_{e}nJ_{n}^{2} & 2k_{x}v_{\perp}J_{n}J_{n}' & -\imath k_{x}v_{z}J_{n}^{2} \\
0 & \imath k_{x}v_{z}J_{n}^{2} & 0 \\
\end{array}\right)\end{equation}
describes the spin evolution leading to the same resonances as the classic evolution $\omega=k_{z}v_{z}+n\Omega_{e}$.
The Bessel functions $J_{n}$ and their derivatives $J_{n}'$ are used here. In this section all Bessel functions $J_{n}$ are functions of $k_{x}v_{\perp}/\Omega_{e}$.

Next, present tensors describing spin evolution manifesting itself at shifted resonances $\omega=k_{z}v_{z}+n\Omega_{e}\pm\Omega_{\mu}$:
\begin{equation}\label{SC_KA Lambda S +}\widehat{\Lambda}_{S+}(n)=\frac{4\pi}{\omega}\frac{\mu_{e}^{2}c^{2}}{2\omega}J_{n}^{2}\biggl(\frac{\partial f_{0}}{\partial\varepsilon}(k_{z}v_{z}+n\Omega_{e})+\frac{2S_{0z}}{\hbar}\biggr)\hat{K},\end{equation}
and
\begin{equation}\label{SC_KA Lambda S -}\widehat{\Lambda}_{S-}(n)=\frac{4\pi}{\omega}\frac{\mu_{e}^{2}c^{2}}{2\omega}J_{n}^{2}\biggl(\frac{\partial f_{0}}{\partial\varepsilon}(k_{z}v_{z}+n\Omega_{e})-\frac{2S_{0z}}{\hbar}\biggr)(\hat{K})^{*},\end{equation}
where
\begin{equation}\label{SC_KA } \hat{K}=\left(
\begin{array}{ccc}
k_{z}^{2} & -\imath k_{z}^{2} & -k_{x}k_{z} \\
\imath k_{z}^{2} & k_{z}^{2} & -\imath k_{x}k_{z} \\
-k_{x}k_{z} & \imath k_{x}k_{z} & k_{x}^{2} \\
\end{array}\right),\end{equation}
and symbol $*$ means the complex conjugation.

The last term in the dielectric permeability tensor (\ref{SC_KA dielectric permeability tensor GF is}) arises as follows
$$\mbox{\boldmath $\Delta$}=\frac{4\pi q_{e}\mu_{e}c}{\omega^{2}}\int\sum_{n=-\infty}^{\infty}\frac{\partial S_{0z}}{\partial\varepsilon}\times$$
\begin{equation}\label{SC_KA DELTA def} \times\{-\imath\Omega_{e}nJ_{n}^{2}, -k_{x}v_{\perp}J_{n}J_{n}', -\imath k_{x}v_{z}J_{n}^{2}\}d\textbf{p}.\end{equation}
The x and z projections of vector $\mbox{\boldmath $\Delta$}$ are equal to zero. The x projection is equal to zero due to explicit summation on $n$ and the z projection is equal to zero due to integration over angles (some details are discussed in the Appendix C). Therefore, the last term in the dielectric permeability tensor (\ref{SC_KA dielectric permeability tensor GF is}) gives contribution in element $\varepsilon^{yy}$ only. This contribution has the following form: $\Delta_{y}\delta^{\alpha y}\delta^{\beta y}$, with $\Delta_{y}=-k_{x}\frac{4\pi q_{e}\mu_{e}c}{\omega^{2}}\int\frac{\partial S_{0z}}{\partial\varepsilon}v_{\perp}J_{0}J_{0}'d\textbf{p}$.

%\begin{equation}\label{SC_KA } \end{equation}

\subsection{Dielectric permeability tensor for the spin-polarized Fermi step distribution function}

The dielectric permeability tensor is obtained above for the general form of isotropic distribution function. In this subsection, a special form of the dielectric permeability tensor is obtained for the spin-1/2 partially polarized 3D electron gas. To this end the equilibrium distribution functions are chosen as follows $f_{0}(p)=[\vartheta(p_{F\uparrow}-p)+\vartheta(p_{F\downarrow}-p)]/(2\pi\hbar)^{3}$ and
$S_{0z}(p)=[\vartheta(p_{F\uparrow}-p)-\vartheta(p_{F\downarrow}-p)]/(2\pi\hbar)^{3}$, where $p_{Fs}=(6\pi^{2}n_{0s})^{\frac{1}{3}}\hbar$.

As it follows from equation (\ref{SCES spin current many part Vector}), the Fermi spin current is related to difference between concentrations of electrons with different spin projections. In kinetic description such difference comes from $S_{0z}$ which is proportional to difference of the Fermi steps of the spin-up and spin-down electrons.

The derivatives of the distribution functions are proportional to the Dirac delta function $\delta(p-p_{Fs})$, since $\partial\vartheta(p_{Fs}-p)/\partial p=-\delta(p-p_{Fs})$. Therefore, the integrals over the module of the momentum module can be easily calculated. As the result we find:
\begin{widetext}$$\varepsilon^{\alpha\beta}=\delta^{\alpha\beta}-\sum_{s=\uparrow,\downarrow}\int \sin\theta d\theta\sum_{n=-\infty}^{\infty}\biggl[\frac{\widetilde{\Lambda}^{\alpha\beta}(n,s)}{\omega-k_{z}v_{Fs}\cos\theta-n\Omega_{e}}
+\frac{mp_{Fs}}{\pi\hbar^{3}}\frac{\mu_{e}^{2}c^{2}}{2\omega^{2}}\sum_{r=+,-} \frac{J_{n}^{2}(k_{z}v_{Fs}\cos\theta+n\Omega_{e})\kappa^{\alpha\beta}_{r}}{\omega-k_{z}v_{Fs}\cos\theta-n\Omega_{e}+r\Omega_{\mu}}$$
\begin{equation}\label{SC_KA dielectric permeability tensor GF} -\frac{1}{\pi\hbar^{3}}\frac{\mu_{e}^{2}c^{2}}{\hbar\omega^{2}}\sum_{r=+,-}\int_{0}^{p_{Fs}} p^{2}dp   \frac{r(-1)^{i_{s}}J_{n}^{2}\kappa^{\alpha\beta}_{r}}{\omega-k_{z}v_{z}-n\Omega_{e}+r\Omega_{\mu}}
-\frac{mp_{Fs}}{\pi\hbar^{3}} \biggl(\mu_{e}^{2}\frac{k_{x}^{2}c^{2}}{\omega^{2}}J_{n}^{2}+\frac{q_{e}\mu_{e}k_{x}c}{\omega^{2}}v_{Fs}\sin\theta J_{n}J_{n}'\biggr)\delta^{\alpha y}\delta^{\beta y}\biggr],\end{equation}
where $\kappa^{\alpha\beta}_{+}=K^{\alpha\beta}$, $\kappa^{\alpha\beta}_{-}=(K^{\alpha\beta})^{*}$, $i_{\uparrow}=0$, $i_{\downarrow}=1$, and
\begin{equation}\label{SC_KA } \widetilde{\Lambda}^{\alpha\beta}(n,s)=\frac{3\omega_{Ls}^{2}}{2\omega v_{Fs}^{2}}\Pi^{\alpha\beta}_{Cl}(n,s) +m^{2}v_{Fs}\biggl(\frac{\mu_{e}^{2}}{\pi\hbar^{3}}\frac{k_{x}^{2}c^{2}}{\omega}J_{n}^{2}\delta^{\alpha y}\delta^{\beta y} +\frac{q_{e}\mu_{e}}{\pi\hbar^{3}}(-1)^{i_{s}}\frac{c}{\omega}\Pi^{\alpha\beta}_{S}(n,s)\biggr),\end{equation}
with
\begin{equation}\label{SC_KA Pi Cl} \widehat{\Pi}_{Cl}(n,s)=\left(\begin{array}{ccc}
\frac{\Omega_{e}^{2}}{k_{x}^{2}}n^{2} J_{n}^{2} & \imath v_{Fs}\sin\theta\frac{\Omega_{e}}{k_{x}} nJ_{n}J_{n}' & v_{Fs}\cos\theta\frac{\Omega_{e}}{k_{x}}n J_{n}^{2} \\
-\imath v_{Fs}\sin\theta\frac{\Omega_{e}}{k_{x}}n J_{n}J_{n}' & v_{Fs}^{2}\sin^{2}\theta(J_{n}')^{2} & -\imath v_{Fs}^{2}\sin\theta\cos\theta J_{n}J_{n}' \\
v_{Fs}\cos\theta \frac{\Omega_{e}}{k_{x}}n J_{n}^{2} & \imath v_{Fs}^{2}\sin\theta\cos\theta J_{n}J_{n}' & v_{Fs}^{2}\cos^{2}\theta J_{n}^{2}
\end{array}\right)\end{equation}\end{widetext}
which has structure similar to well-known from textbooks \cite{Rukhadze book 84}, but it separately describes electrons with spin-up and spin-down,
and the tensor
$$\Pi^{\alpha\beta}_{S}(n,s)=$$
\begin{equation}\label{SC_KA } \left(
\begin{array}{ccc}
0 & \imath\Omega_{e}nJ_{n}^{2} & 0 \\
-\imath\Omega_{e}nJ_{n}^{2} & 2k_{x}v_{Fs}\sin\theta J_{n}J_{n}' & -\imath k_{x}v_{Fs}\cos\theta J_{n}^{2} \\
0 & \imath k_{x}v_{Fs}\cos\theta J_{n}^{2} & 0 \\
\end{array}\right)\end{equation}
describes the spin evolution leading to the same resonances as the classic evolution $\omega=k_{z}v_{Fs}\cos\theta+n\Omega_{e}$.
Here all Bessel functions $J_{n}$ are functions of $k_{x}v_{Fs}\sin\theta/\Omega_{e}$.

\section{Conclusion}

A single fluid spin-1/2 quantum kinetics of electrons has been
applied for a derivation of the dielectric permeability tensor of
spin polarized electron gas. This model consists of two kinetic equations for each species of particles: a generalization of the Vlasov equation containing the spin-spin interaction as an additional term with the spin distribution function and a vector equation for the spin distribution function. Necessity of the application of this model with the spin polarized equilibrium distribution functions follows from the existence of the thermal part of spin current or the Fermi spin current for degenerate electrons which affects properties of transverse waves including the spin-plasma waves. More detailed description of spin current effects requires a kinetic model which has been presented here. Necessary details of the solution of linearized set of kinetic equation required for the derivation of the dielectric permeability tensor have been presented.

\begin{acknowledgements}
The author thanks Professor L. S. Kuz'menkov for fruitful discussions. The work was supported by the Russian
Foundation for Basic Research (grant no. 16-32-00886) and the Dynasty foundation.
\end{acknowledgements}

\section{Appendix A: Linear part of the distribution functions for generalized equilibrium spin distribution functions}

In equilibrium the set of kinetic equations (\ref{SC_KA kinetic equation gen  classic limit with E and B}) and (\ref{SC_KA kinetic equation gen for spin classic limit with E and B}) can be presented in the following form
\begin{equation}\label{SC_KA} \begin{array}{ccc}
\partial_{\varphi}f_{0e\uparrow}=0, & \partial_{\varphi}f_{0e\downarrow}=0, & \partial_{\varphi}f_{0i}=0,
\end{array}\end{equation}
and
\begin{equation}\label{SC_KA eq spin evol eq x} \begin{array}{cc}
\partial_{\varphi}S_{0e,x} =S_{0e,y}, & \partial_{\varphi}S_{0e,y} =-S_{0e,x},
\end{array}\end{equation}
where time and space derivatives of the distribution functions are equal to zero, the equilibrium electric field is equal to zero. The equilibrium magnetic field is equal to the external field: $\textbf{B}_{0}=\textbf{B}_{ext}=B_{0}\textbf{e}_{z}$.

Equations (\ref{SC_KA eq spin evol eq x}) give the general form of dependence of equilibrium spin distribution functions on momentum
$S_{0x}=C(p_{\parallel},p_{\perp})\cos(\varphi+\varphi_{0})$,
$S_{0y}=C(p_{\parallel},p_{\perp})\sin(\varphi+\varphi_{0})$.
Constant $C$ can be equal to zero. It gives the isotropic equilibrium considered in the paper. For isotropic $f_{0}$ and $S_{0z}$ and nonzero constant $C$, functions $S_{0x}$ and $S_{0y}$ can be presented in the following form \cite{Andreev PoP 16 sep kin}:
\begin{equation}\label{SC_KA equilib spin x distrib el un pol} S_{0x}=\frac{1}{(2\pi\hbar)^{3}}\biggl(\Theta(p_{F\uparrow}-p) -\Theta(p_{F\downarrow}-p)\biggr)\cos(\varphi+\varphi_{0}),\end{equation}
\begin{equation}\label{SC_KA equilib spin ydistrib el un pol} S_{0y}=\frac{1}{(2\pi\hbar)^{3}}\biggl(\Theta(p_{F\uparrow}-p) -\Theta(p_{F\downarrow}-p)\biggr)\sin(\varphi+\varphi_{0}).\end{equation}

In equation (\ref{SC_KA kin eq f lin 3D}), there is a change of the last term in the following way: $\delta B_{z}(\textbf{k}\nabla_{\textbf{p}})S_{0z}\rightarrow \delta B_{\beta}(\textbf{k}\nabla_{\textbf{p}})S_{0\beta}=\delta B_{\beta}[(\textbf{k}\textbf{p})\partial_{p}S_{0\beta}/p+\varepsilon^{\beta\gamma z}S_{0\gamma}\varepsilon^{\mu\nu z}k_{\mu}p_{\nu}]$. So, solution (\ref{SC_KA f solution}) is modified to
\begin{widetext}
\begin{equation}\label{SC_KA f solution ext}\delta f=\frac{1}{\Omega_{e}}\int_{C_{0}}^{\varphi}\biggl(q_{e}(\textbf{v}\cdot\delta \textbf{E})\frac{\partial f_{0}}{\partial\varepsilon} +\imath\mu_{e}(\textbf{k}\cdot\textbf{v})\biggl(\delta \textbf{B}\cdot \frac{\partial \textbf{S}_{0}}{\partial\varepsilon}\biggr)
+\imath\mu_{e} ([\delta \textbf{B}, \textbf{S}_{0}]_{z}) \cdot ([\textbf{k},\textbf{p}]_{z})\biggr)
\exp\biggl(\imath\int_{\varphi}^{\varphi'}\frac{(\omega-\textbf{k}\cdot\textbf{v}\mid_{\varphi''})}{\Omega_{e}}d\varphi''\biggr)d\varphi'.\end{equation}

Equation (\ref{SC_KA kin eq S lin 3D}) contains full vector $\textbf{S}_{0}$. So, changing the explicit form of $\textbf{S}_{0}$ we have the spin distribution function evolution equation in the generalized regime.
Equations (\ref{SC_KA Sx lin 3D})-(\ref{SC_KA Sz lin 3D}) are modified to
$$\Omega_{e}\partial_{\varphi}\delta S_{x}+\imath (\omega-\textbf{k}\cdot\textbf{v})\delta S_{x}+\Omega_{\mu}\delta S_{y}$$
\begin{equation}\label{SC_KA Sx lin 3D} =q_{e}\delta \textbf{E}\cdot\nabla_{\textbf{p}}S_{0x}+\imath\mu_{e}(\textbf{k}\cdot\nabla_{\textbf{p}})f_{0}\delta
B_{x}+\frac{2\mu_{e}}{\hbar}S_{0z}\delta B_{y}
-\frac{2\mu_{e}}{\hbar}S_{0y}\delta B_{z} -S_{0y}(mv^{2}\delta B_{z} -mv_{z}(\textbf{v}\delta \textbf{B})),\end{equation}
$$-\Omega_{\mu}\delta S_{x}+\Omega_{e}\partial_{\varphi}\delta S_{y}+\imath (\omega-\textbf{k}\cdot\textbf{v})\delta S_{y}$$
\begin{equation}\label{SC_KA Sy lin 3D}=q_{e}\delta \textbf{E}\cdot\nabla_{\textbf{p}}S_{0y}+\imath\mu_{e}(\textbf{k}\cdot\nabla_{\textbf{p}})f_{0}\delta
B_{y}-\frac{2\mu_{e}}{\hbar}S_{0}\delta B_{x}
+\frac{2\mu_{e}}{\hbar}S_{0x}\delta B_{z} +S_{0x}(mv^{2}\delta B_{z} -mv_{z}(\textbf{v}\delta \textbf{B})), \end{equation}
and
\begin{equation}\label{SC_KA Sz lin 3D} \Omega_{e}\partial_{\varphi}\delta S_{z}+\imath (\omega-\textbf{k}\cdot\textbf{v})\delta S_{z} =q_{e}(\delta \textbf{E}\cdot\nabla_{\textbf{p}})S_{0z}+\imath\mu_{e}(\textbf{k}\cdot\nabla_{\textbf{p}})f_{0}\delta B_{z} +\frac{2\mu_{e}}{\hbar}(\delta B_{x}S_{0y}-\delta B_{y}S_{0x})\end{equation}
(\ref{SC_KA kin eq f lin 3D}) and (\ref{SC_KA kin eq S lin 3D}) to equations (\ref{SC_KA Sx lin 3D})-(\ref{SC_KA Sz lin 3D}).

Therefore, solution (\ref{SC_KA S z solution}) is modified to
\begin{equation}\label{SC_KA S z solution}
\delta S_{z}=\frac{1}{\Omega_{e}}\int_{C_{3}}^{\varphi}\biggl(q_{e}(\textbf{v}\cdot\delta \textbf{E})\frac{\partial S_{0z}}{\partial\varepsilon}
+\imath\mu_{e}(\textbf{k}\cdot\textbf{v})\delta B_{z}\frac{\partial f_{0}}{\partial\varepsilon}
+\frac{2\mu_{e}}{\hbar}(\delta B_{x}S_{0y}-\delta B_{y}S_{0x})\biggr)\exp\biggl(\imath\int_{\varphi}^{\varphi'}\frac{(\omega-\textbf{k}\cdot\textbf{v}\mid_{\varphi''})}{\Omega_{e}}d\varphi''\biggr)d\varphi',\end{equation}

In both regimes solutions for $\delta S_{x}$ and $\delta S_{y}$ are presented by equations () and (), but in the generalized case functions $\Pi_{x}$ and $\Pi_{y}$ have the following form found from equations (\ref{SC_KA Sx lin 3D}) and (\ref{SC_KA Sy lin 3D})
\begin{equation}\label{SC_KA Pi x} \Pi_{x}=\frac{1}{\Omega_{e}}\exp\biggl(\imath\int_{C}^{\varphi}\frac{1}{\Omega_{e}}(\omega-\textbf{k}\cdot\textbf{v}\mid_{\varphi'})d\varphi'\biggr) \biggl(\imath\mu_{e}(\textbf{k}\cdot\nabla_{\textbf{p}})f_{0}\delta
B_{x}+\frac{2\mu_{e}}{\hbar}S_{0z}\delta B_{y} -\frac{2\mu_{e}}{\hbar}S_{0y}\delta B_{z} -S_{0y}(mv^{2}\delta B_{z} -mv_{z}(\textbf{v}\delta \textbf{B}))\biggr),\end{equation}
and
\begin{equation}\label{SC_KA Pi y} \Pi_{y}=\frac{1}{\Omega_{e}}\exp\biggl(\imath\int_{C}^{\varphi}\frac{1}{\Omega_{e}}(\omega-\textbf{k}\cdot\textbf{v}\mid_{\varphi'})d\varphi'\biggr) \biggl(\imath\mu_{e}(\textbf{k}\cdot\nabla_{\textbf{p}})f_{0}\delta
B_{y}-\frac{2\mu_{e}}{\hbar}S_{0}\delta B_{x} +\frac{2\mu_{e}}{\hbar}S_{0x}\delta B_{z} +S_{0x}(mv^{2}\delta B_{z} -mv_{z}(\textbf{v}\delta \textbf{B}))\biggr).\end{equation}
\end{widetext}

\section{Appendix B: An intermediate form of the conductivity tensor}

More explicit form of the conductivity tensor caused by the current is obtained at substituting of the distribution function $\delta f$
$$\sigma^{\alpha\beta}_{1}(\omega)=\frac{q_{e}}{m}\frac{1}{\Omega_{e}}\int d\textbf{p}p^{\alpha} \int_{C_{0}}^{\varphi}d\varphi'\biggl(q_{e}v^{\beta}\mid_{\varphi'}\frac{\partial f_{0}}{\partial\varepsilon} $$
\begin{equation}\label{SC_KA sigma 1 expl} +\imath\mu_{e}(\textbf{k}\cdot\textbf{v})\mid_{\varphi'}\frac{k_{x}c}{\omega}\delta^{y\beta}\frac{\partial S_{0z}}{\partial\varepsilon}\biggr)
\exp\biggl(\imath\int_{\varphi}^{\varphi'}\frac{\omega-\textbf{k}\cdot\textbf{v}\mid_{\varphi''}}{\Omega_{e}}d\varphi''\biggr).\end{equation}

The conductivity tensor caused by the curl of magnetization has a complicate structure. Hence, it is splitted on three parts:
\begin{equation}\label{SC_KA sigma 2 x}\sigma^{x\beta}_{2}(\omega)\delta E_{\beta}=-\imath\mu_{e}c\int k_{z}\delta S_{y}d\textbf{p},\end{equation}
\begin{equation}\label{SC_KA sigma 2 y}\sigma^{y\beta}_{2}(\omega)\delta E_{\beta}=\imath\mu_{e}c\int(k_{z}\delta S_{x}-k_{x}\delta S_{z})d\textbf{p},\end{equation}
and
\begin{equation}\label{SC_KA sigma 2 z}\sigma^{z\beta}_{2}(\omega)\delta E_{\beta}=\imath\mu_{e}c\int k_{x}\delta S_{y} d\textbf{p}.\end{equation}

Explicit forms of functions (\ref{SC_KA sigma 2 x})-(\ref{SC_KA sigma 2 z}) appear at substitution of the distribution functions $\delta S_{x}$, $\delta S_{y}$, and $\delta S_{z}$. Thus, the following expression can be found for $\sigma^{x\beta}_{2}(\omega)$:
\begin{widetext}
$$\sigma^{x\beta}_{2}(\omega)\delta E_{\beta}=-\mu_{e}^{2}\frac{k_{z}c^{2}}{2\Omega_{e}\omega}
\int\Biggl\{\int_{C_{1}}^{\varphi}\exp\biggl(\frac{\imath\Omega_{\mu}}{\Omega_{e}}(\varphi-\varphi')\biggr)
\exp\biggl(\imath\int_{\varphi}^{\varphi'}\frac{\omega-\textbf{k}\cdot\textbf{v}\mid_{\varphi''}}{\Omega_{e}}d\varphi''\biggr)
\Biggl[\biggl(-(\imath\textbf{k}\cdot\textbf{v}\mid_{\varphi'})\frac{\partial f_{0}}{\partial\varepsilon} +\imath\frac{2S_{0z}}{\hbar}\biggr)k_{z}\delta E_{y}$$
$$+\biggl(\imath(\imath\textbf{k}\cdot\textbf{v}\mid_{\varphi'})\frac{\partial f_{0}}{\partial\varepsilon} +\frac{2S_{0z}}{\hbar}\biggr)(k_{z}\delta E_{x}-k_{x}\delta E_{z})\Biggr]d\varphi'
+ \int_{C_{2}}^{\varphi} \exp\biggl(\frac{\imath\Omega_{\mu}}{\Omega_{e}}(\varphi'-\varphi)\biggr) \exp\biggl(\imath\int_{\varphi}^{\varphi'}\frac{\omega-\textbf{k}\cdot\textbf{v}\mid_{\varphi''}}{\Omega_{e}}d\varphi''\biggr)\times $$
\begin{equation}\label{SC_KA } \times\Biggl[\biggl((\imath\textbf{k}\cdot\textbf{v}\mid_{\varphi'})\frac{\partial f_{0}}{\partial\varepsilon}
+\imath\frac{2S_{0z}}{\hbar}\biggr)k_{z}\delta E_{y}+\biggl(\imath(\imath\textbf{k}\cdot\textbf{v}\mid_{\varphi'})\frac{\partial f_{0}}{\partial\varepsilon} -\frac{2S_{0z}}{\hbar}\biggr)(k_{z}\delta E_{x}-k_{x}\delta E_{z})\Biggr]d\varphi'\Biggr\}d\textbf{p}.\end{equation}
Neglecting the anomalous magnetic moment of electron we have $\Omega_{\mu}=\Omega_{e}$. Hence, $\sigma^{x\beta}_{2}(\omega)$ consists of two parts proportional to $e^{\imath(\varphi-\varphi')}$ or $e^{-\imath(\varphi-\varphi')}$.
Elements $\sigma^{y\beta}_{2}(\omega)$ have the following explicit form:
$$\sigma^{y\beta}_{2}(\omega)\delta E_{\beta}=\imath\mu_{e}^{2}\frac{k_{z}c^{2}}{2\Omega_{e}\omega}
\int\Biggl\{\int_{C_{1}}^{\varphi}\exp\biggl(\frac{\imath\Omega_{\mu}}{\Omega_{e}}(\varphi-\varphi')\biggr)
\exp\biggl(\imath\int_{\varphi}^{\varphi'}\frac{\omega-\textbf{k}\cdot\textbf{v}\mid_{\varphi''}}{\Omega_{e}}d\varphi''\biggr)
\Biggl[\biggl(-(\imath\textbf{k}\cdot\textbf{v}\mid_{\varphi'})\frac{\partial f_{0}}{\partial\varepsilon} +\imath\frac{2S_{0z}}{\hbar}\biggr)k_{z}\delta E_{y}$$
$$+\biggl(\imath(\imath\textbf{k}\cdot\textbf{v}\mid_{\varphi'})\frac{\partial f_{0}}{\partial\varepsilon} +\frac{2S_{0z}}{\hbar}\biggr)(k_{z}\delta E_{x}-k_{x}\delta E_{z})\Biggr]d\varphi'
+\int_{C_{2}}^{\varphi}\exp\biggl(\frac{\imath\Omega_{\mu}}{\Omega_{e}}(\varphi'-\varphi)\biggr) \exp\biggl(\imath\int_{\varphi}^{\varphi'}\frac{\omega-\textbf{k}\cdot\textbf{v}\mid_{\varphi''}}{\Omega_{e}}d\varphi''\biggr)\times$$
$$ \times\Biggl[\biggl(-(\imath\textbf{k}\cdot\textbf{v}\mid_{\varphi'})\frac{\partial f_{0}}{\partial\varepsilon} -\imath\frac{2S_{0z}}{\hbar}\biggr)k_{z}\delta E_{y}
+\biggl(-\imath(\imath\textbf{k}\cdot\textbf{v}\mid_{\varphi'})\frac{\partial f_{0}}{\partial\varepsilon} +\frac{2S_{0z}}{\hbar}\biggr)(k_{z}\delta E_{x}-k_{x}\delta E_{z})\Biggr]d\varphi'$$
\begin{equation}\label{SC_KA }
-k_{x}\int_{C_{3}}^{\varphi}\biggl(\frac{q_{e}}{\mu_{e}}(\textbf{v}\cdot\delta \textbf{E})\mid_{\varphi'}\frac{\partial S_{0z}}{\partial\varepsilon}+\imath(\textbf{k}\cdot\textbf{v})\mid_{\varphi'}\frac{k_{x}c}{\omega}\delta E_{y}\frac{\partial f_{0}}{\partial\varepsilon}\biggr) \exp\biggl(\imath\int_{\varphi}^{\varphi'}\frac{\omega-\textbf{k}\cdot\textbf{v}\mid_{\varphi''}}{\Omega_{e}}d\varphi''\biggr)d\varphi'
\Biggr\}d\textbf{p}.\end{equation}\end{widetext}
Similar to $\sigma^{x\beta}_{2}(\omega)$, $\sigma^{y\beta}_{2}(\omega)$ consists of two parts proportional to $e^{\imath(\varphi-\varphi')}$ or $e^{-\imath(\varphi-\varphi')}$.
Elements $\sigma^{z\beta}_{2}$ can be simply presented via $\sigma^{x\beta}_{2}$
\begin{equation}\label{SC_KA sigma z expl} \sigma^{z\beta}_{2}(\omega)=-\frac{k_{x}}{k_{z}}\sigma^{x\beta}_{2}(\omega)\end{equation}
in accordance with equations (\ref{SC_KA sigma 2 x}) and (\ref{SC_KA sigma 2 z}).

\section{Appendix C: Calculation of several elements in dielectric permeability tensor}

For calculation of z-projection of $\mbox{\boldmath $\Delta$}$ (\ref{SC_KA DELTA def})  we need to consider integral over $\theta$. Dependence on $\theta$ comes from $v_{z}J_{n}^{2}(v_{\perp})$. So, we have integral $I_{n}=\int\sin\theta d\theta \cos\theta J_{n}^{2}(a\sin\theta)$, where $a\equiv k_{x}v_{Fs}/\Omega_{e}$. Using $J_{-n}(z)=(-1)^{n}J_{n}(z)$ and $J_{-n}^{2}(z)=J_{n}^{2}(z)$ we reduce our calculations to nonnegative Bessel functions $n\geq0$. Explicit integration can be performed at the application of series expansion of the Bessel function:
\begin{equation}\label{SC_KA } J_{n}(z)=\sum_{k=0}^{\infty}\frac{(-1)^{k}z^{n}(z/2)^{2k}}{2^{n}k!\Gamma(n+k+1)}.\end{equation}
Integrating we obtain that $I_{n}\sim\sin^{2k+2l+2n+2}\theta\mid_{0}^{\pi}=0$, where $k,l,n\geq0$ and $l$ is used in expansion of the second Bessel function presented under the integral.
x-projection of $\mbox{\boldmath $\Delta$}$ is proportional to $nJ_{n}^{2}$ we can explicitly calculate sum of these terms and find that it is equal to zero:
$$\sum_{n=-\infty}^{\infty}nJ_{n}^{2}=\sum_{n=1}^{\infty}nJ_{n}^{2}+\sum_{n=-1}^{-\infty}nJ_{n}^{2}$$
\begin{equation}\label{SC_KA } =\sum_{n=1}^{\infty}(nJ_{n}^{2}+(-n)J_{-n}^{2})=0,\end{equation}
where we have used $J_{-n}(z)=(-1)^{n}J_{n}(z)$.
Similarly, applying equation $J_{n}'(z)=J_{n-1}(z)-\frac{n}{z}J_{n}(z)$, we find $\sum_{n=-\infty}^{\infty}J_{n}J_{n}'=J_{0}J_{0}'$. It simplifies y-projection of $\mbox{\boldmath $\Delta$}$.
%\begin{equation}\label{SC_KA } \end{equation}

\end{document}